\documentclass[graybox, envcountchap]{svmult}

\usepackage{mathptmx}        % selects Times Roman as basic font
\usepackage{amsmath}
\usepackage{amssymb}
\usepackage{color}
\usepackage{helvet}          % selects Helvetica as sans-serif font
\usepackage{courier}         % selects Courier as typewriter font
\usepackage{dirtree}
\usepackage{makeidx}        % allows index generation
\usepackage{graphicx}        % standard LaTeX graphics tool
\usepackage{subfig}

\usepackage{multicol}        % used for the two-column index
\usepackage[bottom]{footmisc}% places footnotes at page bottom

\usepackage{hyperref}        %for hyperlinks
\hypersetup{colorlinks=true,urlcolor=blue}

\usepackage[misc]{ifsym}
\usepackage{cite}

\makeindex             % used for the subject index
\usepackage{soul}

\begin{document}

%%%%%%%%%%%%%%%%%%%%%%%%%%%%%%%%%%%%%%%%%%%%%%%%%%%%%%%%%%%%%%%%%

\title{Decaying Dark Matter and the Hubble Tension}
\author{Andreas Nygaard, Emil Brinch Holm, Thomas Tram, and Steen Hannestad}
\institute{Andreas Nygaard (\Letter) \at Department of Physics and Astronomy, Aarhus University, DK-8000 Aarhus C, Denmark, \email{andreas@phys.au.dk}
\and Emil Brinch Holm \at Department of Physics and Astronomy, Aarhus University, DK-8000 Aarhus C, Denmark, \email{ebholm@phys.au.dk}
\and Thomas Tram \at Department of Physics and Astronomy, Aarhus University, DK-8000 Aarhus C, Denmark, \email{thomas.tram@phys.au.dk}
\and Steen Hannestad \at Department of Physics and Astronomy, Aarhus University, DK-8000 Aarhus C, Denmark, \email{steen@phys.au.dk}}
\maketitle
\vspace{-0.5in}
\abstract{Decaying dark matter models generically modify the equation of state around the time of dark matter decay, and this in turn modifies the expansion rate of the Universe through the Friedmann equation. Thus, a priori, these models could solve or alleviate the Hubble tension, and depending on the lifetime of the dark matter, they can be classified as belonging to either the early- or late-time solutions.
Moreover, decaying dark matter models can often be realized in particle physics models relatively easily. However, the implementations of these models in Einstein--Boltzmann solver codes are non-trivial, so not all incarnations have been tested. 
It is well known that models with very late decay of dark matter do not alleviate the Hubble tension, and in fact, cosmological data puts severe constraints on the lifetime of such dark matter scenarios.
However, models in which a fraction of the dark matter decays to dark radiation at early times hold the possibility of modifying the effective equation of state around matter-radiation equality without affecting late-time cosmology. This scenario is therefore a simple realization of a possible early-time solution to the Hubble tension,
and cosmological parameter estimation with current data in these models yields a value of $H_0 = 68.73^{+0.81}_{-1.3}$ at $68\%$ C.I.~\cite{Holm:2022eqq}. This still leads to a $2.7\sigma$ Gaussian tension with the representative local value of $H_0 = 73.2 \pm 1.3$ km s$^{-1}$ Mpc$^{-1}$~\cite{Riess:2020fzl}. Additional work is, however, required to test more complex decay scenarios, which could potentially prefer higher values of $H_0$ and provide a better solution to the Hubble tension.} 
% Steen's modified abstract 23/6 - 248 words

%First go at abstract, 157 words so far. Needs to be expanded a bit. (And/or rewritten!)
%Each chapter should be preceded by an abstract (about 250 words) that summarizes the content. The abstract will also appear \textit{online} at \url{www.SpringerLink.com} and be available with unrestricted access. This allows unregistered users to read the abstract as a teaser for the complete chapter.

%%%%%%%%%%%%%%%%%%%%%%%%%%%%%%%%%%%%%%%%%%%%%%%%%%%%%%%%%
\newpage

\section{Introduction}
\noindent Although the nature of dark matter remains unknown, a brief look at the Standard Model contents of the Universe reveals that a majority of the known particles are unstable and decay. By analogy, a natural question to ask is whether dark matter may decay on cosmological timescales. Decays of dark matter into electromagnetically interacting particles are strongly constrained by CMB observations~\cite{Zhang:2007zzh}. Decays into a dark sector, so-called \textit{invisible decays}, on the other hand, are much less constrained because no direct observation channel exists. Nonetheless, there are strong constraints on models that assume \textit{all} of dark matter to decay on cosmological timescales (e.g., the simple observation that we observe it today)~\cite{Audren:2014bca,Simon:2022ftd}. However, these constraints may always be evaded by considering a scenario where only a fraction of the dark matter decays invisibly. It is this class of models we study in this chapter.

There exist several phenomenological models of invisibly decaying dark matter, largely varying in their assumptions on the decaying particle (cold or warm) and on the decay products (massive or massless, two- or many-body decays). In this chapter, we review constraints on the three most studied models: 
\begin{description}
	\item \textbf{DCDM$\rightarrow$DR}: Decaying cold dark matter (DCDM) decaying into dark radiation (DR). Presented in section~\ref{sec:dcdm_dr}.
	\item \textbf{DCDM$\rightarrow$DR+WDM}: Decaying cold dark matter decaying into warm dark matter (WDM) \textit{and} dark radiation. Presented in section~\ref{sec:dcdm_wdmdr}.
	\item \textbf{DWDM$\rightarrow$DR}: Decaying \textit{warm} dark matter (DWDM) decaying into dark radiation. Presented in section~\ref{sec:dwdm_dr}.
\end{description}

\begin{figure}[b]
	\includegraphics[width=\textwidth]{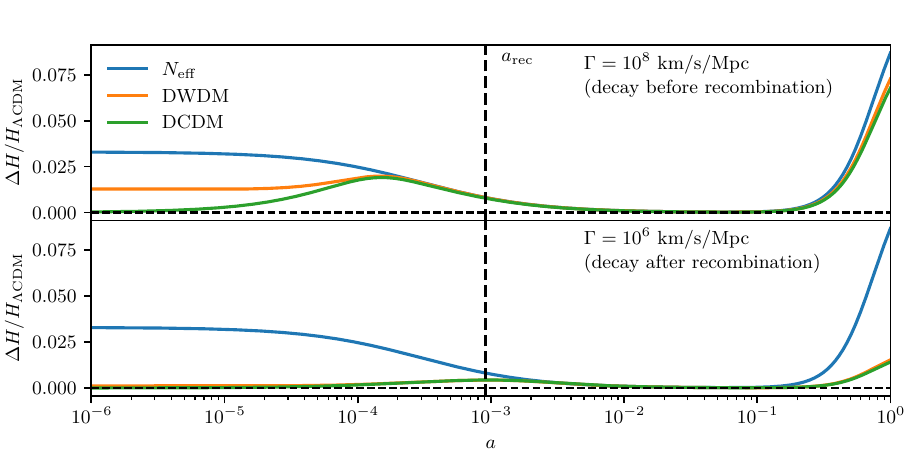}
	\caption{\textsl{Hubble parameter $H$ as a function of scale factor $a$, relative to its value in the $\Lambda$CDM model, for models of decaying cold dark matter (DCDM), decaying warm dark matter (DWDM), and additional relativistic degrees of freedom, $\Delta N_\mathrm{eff}$. Taken from Ref.~\cite{Holm:2022eqq}.}}
	\label{fig:h0}
\end{figure}

\noindent Why are these models relevant to the Hubble tension? Figure~\ref{fig:h0} shows the relative change in the Hubble parameter $H(a)$ as a function of the scale factor $a$ for the DWDM$\rightarrow$DR and DCDM$\rightarrow$DR models, as well as a model with additional relativistic degrees of freedom, $\Delta N_\mathrm{eff}$. The abundances of the species have been fixed such that they all contribute $\Delta N_\mathrm{eff}=0.5$ to the number of relativistic degrees of freedom at $a=1$, the acoustic scale is fixed (thus letting $H_0$ vary) to simulate observational constraints, and the DWDM and DCDM models have the same decay rate $\Gamma$. 

In the top panel, $\Gamma$ is such that both models decay before recombination. In this case, the energy density is greater than that of $\Lambda$CDM, increasing the value of the Hubble parameter. As the species decay, their values of $H$ converge to that of the $N_\mathrm{eff}$ model. In all three models, the increase in $H(a)$ prior to recombination decreases the sound horizon at recombination, $r_s (a_*)$. Since observations essentially fix the acoustic scale to $\theta_s = r_s (a_*)/ D_A (a_*)$, where $D_A(a_*)$ is the angular diameter distance to recombination, this results in a decrease of $D_A (a_*)$, which manifests in an increased value of $H_0$, as can be seen in the figure\footnote{However, in the bottom panel, where $\Gamma$ is such that both models decay after recombination, only a comparatively negligible increase in $H_0$ is seen.}. The model of extra relativistic degrees of freedom obtains the largest increase in $H_0$ but is known to have too strong an impact on the CMB spectrum to satisfactorily solve the Hubble tension~\cite{Schoneberg:2021qvd}. The motivation for the decaying models, then, is that they may be able to circumvent these constraints by virtue of injecting the radiation energy density more locally around recombination. 

%%%%%%%%%%%%%%%%%%%%%%%%%%%%%%%%%%%%%%%%%%%%%%%%%%%%%%%%%

\section{DCDM$\rightarrow$DR} \label{sec:dcdm_dr}

The most simple model involving decaying dark matter is decaying \textit{cold} dark matter (DCDM), where a cold mother particle decays into massless particles denoted as \textit{dark radiation} (DR), which is a type of radiation that does not interact electromagnetically. We can simply write this process as
\begin{align}\nonumber
X_{\rm dcdm} \rightarrow \gamma_{\rm dr}\, .
\end{align}
As mentioned in the introduction, the scenario where all dark matter is decaying is heavily constrained, so we consider a model where just a fraction,
\begin{align}
f_{\rm dcdm} \equiv \frac{\Omega_{\rm dcdm}^{\rm ini}}{\Omega_{\rm dcdm}^{\rm ini} + \Omega_{\rm scdm}}\, ,
\end{align}
of the dark matter is unstable. Here $\Omega_{\rm scdm}$ is the density parameter of the stable cold dark matter, and $\Omega_{\rm dcdm}^{\rm ini}$ is the density parameter of the decaying component as it would be today if none of it had decayed~\cite{Nygaard:2020sow}. This partial decay can also be interpreted as all CDM decaying into a stable CDM particle and DR~\cite{Nygaard:2020sow}.\\

\noindent The equations for the homogeneous and isotropic energy densities in this model resemble the normal background equations for CDM and radiation but with additional source terms dependent on the decay rate, $\Gamma_{\rm dcdm}$ (w.r.t. proper time)~\cite{Nygaard:2020sow},
\begin{equation}\label{eq:dcdm_fluid}
\begin{aligned}
    \rho'_{\mathrm{dcdm}}&=-3\frac{a'}{a}\rho_{\mathrm{dcdm}}-a\Gamma_{\mathrm{dcdm}}\rho_{\mathrm{dcdm}}\,,\\
    \rho'_{\mathrm{dr}}&=-4\frac{a'}{a}\rho_{\mathrm{dr}}+a\Gamma_{\mathrm{dcdm}}\rho_{\mathrm{dcdm}}\,,
\end{aligned}
\end{equation}
where the prime denotes derivatives w.r.t. conformal time.

The equations in first-order general relativistic perturbation theory for the cold mother particle are exactly the same as for normal CDM, where only the density perturbation is non-trivial. For DR, the Boltzmann hierarchy resembles that of massless neutrinos but with additional source terms in the lowest modes~\cite{Audren:2014bca}.\\

\begin{figure}[t]
	\makebox[\textwidth][c]{\includegraphics[width=\textwidth]{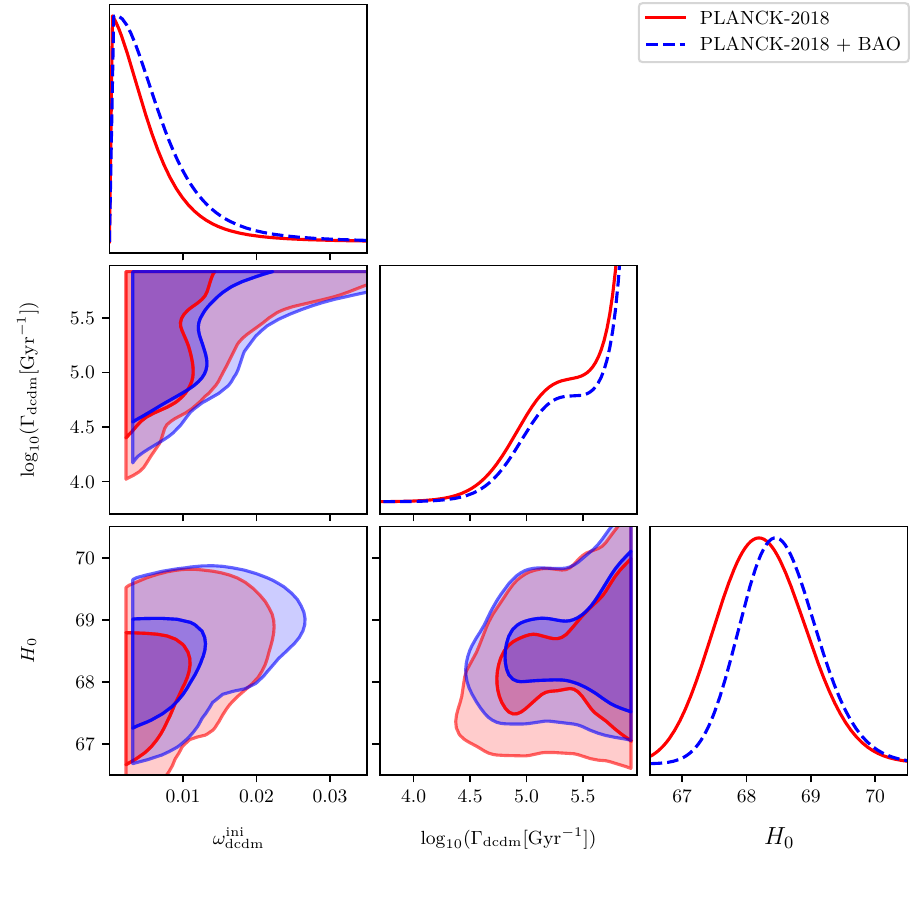}}
	\caption{\textsl{Triangle plot of posteriors in the short-lived regime using Planck 2018 data with and BAO peak data. The figure shows the model-specific parameters, $\omega^{\rm ini}_{\rm dcdm}$ and $\Gamma_{\rm dcdm}$, along with $H_0$, and has been produced using the same MCMC runs used in Ref.~\cite{Nygaard:2020sow}.}}
	\label{fig:dcdm_posterior}
\end{figure}

\begin{figure}
\centering
\begin{minipage}[t]{.48\textwidth}
  \centering
  \includegraphics[width=\textwidth]{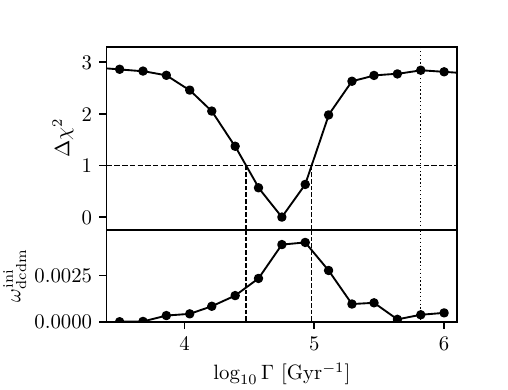}
	\caption{\textsl{Top panel: One-dimensional profile likelihood of $\log_{10} \Gamma/$Gyr. Bottom panel: The abundance of decaying cold dark matter $\omega^{\mathrm{ini}}_\mathrm{dcdm}$ associated with every point in the profile above. Dashed lines indicate the $\Delta \chi^2 = 1$ intersections, giving the $1\sigma$ CIs. Taken from Ref.~\cite{Holm:2022kkd}.}}
	\label{fig:dcdm_profile}\end{minipage}%
\hspace{0.034\textwidth}
\begin{minipage}[t]{.48\textwidth}
  \centering
  \includegraphics[width=\textwidth]{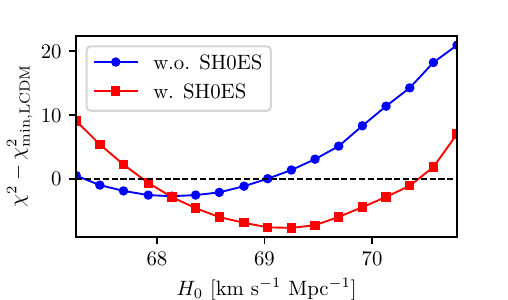}
	\caption{\textsl{One-dimensional profile likelihoods of $H_0$ using Planck 2018 data and the full BOSS DR12 data set (see text). The two profiles are with and without an additional S$H0$ES prior. Taken from Ref.~\cite{Holm:2022kkd}.}}
	\label{fig:H0_profile}
\end{minipage}
\end{figure}

\noindent It is useful to distinguish between two regions of the parameter space when dealing with this model, since the two limits $\Gamma_{\rm dcdm}\rightarrow 0$ and $\Gamma_{\rm dcdm}\rightarrow \infty$ correspond to the $\Lambda$CDM model, with the first having no constraints on the fractional amount of DCDM, $f_{\rm dcdm}$, and the latter having a very weak constraint of $f_{\rm dcdm}\neq 1$. This leads to a large parameter space volume of good fits to data in both limiting cases, and this funnel-shaped likelihood is difficult to sample simultaneously with an MCMC algorithm. We therefore split the parameter space into a \textit{long-lived} region, where $\Gamma_{\rm dcdm}<10^3\; {\rm Gyr}^{-1}$, and a \textit{short-lived} region, where $\Gamma_{\rm dcdm}>10^3\; {\rm Gyr}^{-1}$. The long-lived regime corresponds to a very late decay long after recombination and the short-lived regime corresponds to an early decay before recombination. The latter is probably the most interesting in relation to the Hubble tension, since it can impact the physics around recombination, and hence the formation of the CMB. The constraints on $H_0$ from Ref.~\cite{Nygaard:2020sow} are different for the two regimes, with the long-lived regime yielding a 95.45\% credible interval of $H_0 = 67.7^{+1.0}_{-0.9}\;{\rm km}\; {\rm s}^{-1}\; {\rm Mpc}^{-1}$ and the short-lived one yielding one of $H_0 = 68.6^{+1.2}_{-1.4}\;{\rm km}\; {\rm s}^{-1}\; {\rm Mpc}^{-1}$. As we might expect, the short-lived regime alleviates the Hubble tension better of the two with almost one standard deviation (compared to a value inferred from local measurements of $H_0 = 73.2 \pm 1.3 \; {\rm km}\; {\rm s}^{-1}\; {\rm Mpc}^{-1}$ from Ref.~\cite{Riess:2020fzl}). We will therefore only consider the short-lived regime from hereon.

The DCDM model has been investigated in numerous papers with Bayesian statistics~\cite{Ichiki:2004vi,Berezhiani:2015yta,Chudaykin:2016yfk,Poulin:2016nat,Oldengott:2016yjc,Chudaykin:2017ptd,Pandey:2019plg,Xiao:2019ccl} and recently also with frequentist statistics and profile likelihoods~\cite{Holm:2022kkd}. Figure \ref{fig:dcdm_posterior} shows the posteriors for the short-lived regime in the model-specific parameters (along with $H_0$) using high-$\ell$ and low-$\ell$ temperature and polarization as well as lensing data from Planck 2018~\cite{Planck:2018vyg} (from now on just referred to as Planck 2018 data) with and without the BOSS DR12 BAO peak data~\cite{BOSS:2016wmc} (from now on referred to as BAO peak data). The posterior of the decay rate, $\Gamma_{\rm dcdm}$, shows a volume effect towards higher values, where a large part of the parameter space fits the data well. Because of this, it is not possible to assign a reasonable credible interval to the decay rate since any such interval would be prior dependent. It is, however, also apparent that a region around ${\rm log}_{10}(\Gamma / {\rm Gyr})\in[4.5, 5.5]$ includes a separate effect in the posterior, visualized as a small "bump" or plateau, which marks this as a region of interest. By investigating the same region in the decay rate with profile likelihoods, we find that a peak in the likelihood (a "well" in $\chi^2 (\Gamma) \equiv -2 \log(\mathcal{L} (\Gamma)/\max (\mathcal{L}(\Gamma))$) arises here, as shown in figure \ref{fig:dcdm_profile}. This study has been performed with the same data (Planck 2018 data and BAO peak data) along with low redshift BAO data from the 6dF survey~\cite{Beutler:2011hx} and the BOSS main galaxy sample~\cite{Ross:2014qpa} (the inclusion of which will be referred to as the full BOSS DR12 data set). The approximate $68\%$ confidence interval obtained is $\log_{10} \Gamma \text{ Gyr}^{-1} = 4.763^{+0.214}_{-0.290}$, while it is unconstrained at $95 \%$. The global best-fit of the model lies in this region at $\Delta \chi^2 = 2.8$ relative to the $\Lambda$CDM model, hinting at a mild preference for the DCDM model over $\Lambda$CDM. Interestingly, the best-fitting parameters $\log_{10} \Gamma \text{ Gyr}^{-1}=4.763$ and $\omega^{\rm ini}_{\rm dcdm}=0.00429$ correspond to a scenario where about $3\%$ of the cold dark matter decays just prior to recombination, in support of the tendency of early time solutions to the Hubble tension to preferentially modify physics temporally close to recombination~\cite{Knox:2019rjx, Schoneberg:2021qvd}. Despite the stronger signature of DCDM in the frequentist analysis, the resulting $68\%$ constraint $H_0=68.14^{+0.54}_{-0.49}$ km s$^{-1}$ Mpc$^{-1}$ solidifies the conclusion that the DCDM$\rightarrow$DR model is unable to solve the $H_0$ tension.

%%%%%%%%%%%%%%%%%%%%%%%%%

\section{DCDM$\rightarrow$DR+WDM} \label{sec:dcdm_wdmdr}

A reasonable increase in model complexity is to allow one of the daughter particles to be massive. We still have a cold mother particle decaying into dark radiation, but now also accompanied by a massive daughter particle acting as \textit{warm dark matter} (WDM). We can write this process as
\begin{align}\nonumber
X_{\rm dcdm} \rightarrow \gamma_{\rm dr} + Y_{\rm wdm}\, .
\end{align}

\noindent This model has the same model parameters as the DCDM$\rightarrow$DR model (abundance and decay rate), but it also has a new parameter: The mass ratio of the WDM particle to the DCDM particle, $\widetilde{m}$. Using conservation of energy and momentum, one can relate this mass ratio to the WDM velocity and the fraction of energy transferred to the massless DR particle in the decay, $\epsilon$, through the following two equations~\cite{Blackadder:2014wpa},
\begin{align}
	\epsilon = \frac{1}{2}(1-\widetilde{m}^2),\qquad\qquad \beta_{\rm wdm}^2 = \frac{\epsilon^2}{(1-\epsilon)^2}\, ,
\end{align}
where $\beta_{\rm wdm}$ is the velocity of the massive daughter in natural units.

The background and perturbation equations are the same for DCDM and DR as in the DCDM$\rightarrow$DR model, except for an additional factor of $\epsilon$ on the source term in the background equation for DR. The equations for the massive daughter can be expressed in terms of the equation-of-state parameter, $w_{\rm wdm}(a)$, which can be shown to have the following form~\cite{Blackadder:2014wpa},
\begin{align}
    w_{\rm wdm}(a) = \frac{1}{3} \, \frac{\Gamma \beta^2}{1-\mathrm{e}^{-\Gamma t}} \int_{a_{\rm ini}}^{a(t)} \frac{\mathrm{e}^{-\Gamma t_D}\; \mathrm{d}ln(a_D)}{H_D[(a/a_D)^2(1-\beta^2)+\beta^2]},
\end{align}
where $a_{\rm ini}$ refers to some initial value of $a$ where our numerical integration begins, and the subscript "$D$" refers to the integration variable. The background equation for the massive daughter particle is then
\begin{align}
    \dot{\rho}_{\rm wdm}&=-3\big(1+w_{\rm wdm}(a)\big)\frac{a'}{a}\rho_{\rm wdm}+(1-\epsilon)a\Gamma\rho_0\,.
\end{align}
The perturbation equations resemble those of massive neutrinos but can be rewritten in terms of $w_{\rm wdm}$ in a similar manner~\cite{FrancoAbellan:2021sxk}. An accurate implementation of the full Boltzmann hierarchy fast enough for MCMC runs to be feasible is still missing, but results have been produced using a fluid approximation~\cite{FrancoAbellan:2021sxk, FrancoAbellan:2020xnr, Schoneberg:2021qvd, Simon:2022ftd, Fuss:2022zyt}.

\begin{figure}[t]
	\includegraphics[width=\textwidth]{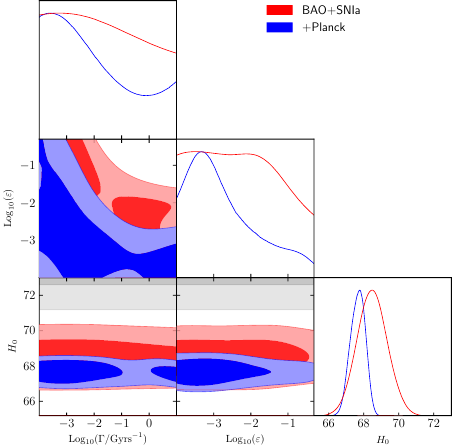}
	\caption{\textsl{Triangle plot of the decay rate, the energy transfer ratio, and the Hubble constant. The red lines and contours are from a background-only MCMC run with the full BOSS DR12 data set and a S$H0$ES prior, while the blue lines and contours are from an MCMC run including perturbations (with the fluid approximation) and also the Planck 2018 data in addition to the other data sets. The gray contours are the $H_0$ measurement by the S$H0$ES collaboration. The figure is taken from Ref.~\cite{FrancoAbellan:2021sxk}.}}
	\label{fig:abellan}
\end{figure}

Figure \ref{fig:abellan} shows the posteriors for the model-specific parameters, $\Gamma$ and $\epsilon$, along with $H_0$. These results assume that all dark matter is decaying, i.e., the abundance ratio, $f_{\rm dcdm}$, is fixed to unity. Because of this, the results are within the very long-lived regime, and the energy transfer ratio, $\epsilon$, is rather low as well, resulting in a heavy WDM daughter particle acting much like CDM. In particular, if ${\rm log}_{10}(\epsilon)\lesssim-2.7$, the decay rate becomes unconstrained because the massive daughter particle becomes virtually indistinguishable from the cold mother particle. The 68.27\% credible interval for the Hubble constant from this analysis is $H_0 = 67.71^{+0.42}_{-0.43}\; {\rm km}\; {\rm s}^{-1}\; {\rm Mpc}^{-1}$, which is similar to the result from the long-lived regime in the simple DCDM$\rightarrow$DR model. We would expect that an analysis with a fractional decay in the short-lived regime would yield a higher value of $H_0$. In order to do this accurately, the full hierarchy should be solved for WDM instead of using a fluid approximation.

%%%%%%%%%%%%%%%%%%%%%%%%%

\section{DWDM$\rightarrow$DR} \label{sec:dwdm_dr}

Another extension of the simplest decaying dark matter model in section~\ref{sec:dcdm_dr} is to allow the decaying particle a non-negligible momentum, making it a warm dark matter species. The model of a decaying warm dark matter (DWDM) species decaying to dark radiation was studied in Refs.~\cite{Holm:2022eqq, Blinov:2020uvz}. This model has several interesting particle physics realizations, such as decaying neutrinos~\cite{Barenboim:2020vrr} and majoron decays~\cite{Escudero:2021rfi}. In particular, in~\cite{FrancoAbellan:2021hdb} it was used to constrain the lifetime of the active neutrino species.

The fundamental characteristic that separates the DWDM model from the other decaying dark matter models is that the non-negligible momentum increases the lifetime by a factor of $E/m$ through time dilation, where $E$ and $m$ denote the energy and mass of the decaying particle, respectively. In the general case, the model contains three parameters: The decay constant $\Gamma$ (or, equivalently, the lifetime $\tau \equiv 1/\Gamma$), the DWDM mass $m$ and the abundance of the DWDM species, specified either at final or initial time. The background equations are~\cite{Holm:2022eqq}

\begin{equation}\label{eq:dwdm_fluid}
	\begin{aligned}
		\rho'_{\mathrm{dwdm}}&=-3\frac{a'}{a}(\rho_{\mathrm{dwdm}} + p_{\mathrm{dwdm}} ) - a\Gamma m n_\mathrm{dwdm},\\
		\rho'_{\mathrm{dr}}&=-4\frac{a'}{a}\rho_{\mathrm{dr}}+a\Gamma_{\mathrm{dcdm}}m n_\mathrm{dwdm},
	\end{aligned}
\end{equation}
where $\rho_i$ and $p_i$ denote the energy and pressure density of the $i$'th species and $n_\mathrm{dwdm}$ the number density of the decaying species. Apart from the addition of the non-zero DWDM pressure $p_\mathrm{dwdm}$, these equations are identical to the case of a cold decaying species~\eqref{eq:dcdm_fluid} up to the substitution $\rho \rightarrow mn$, i.e., it is the rest mass and not the total energy that drives the decay.

As in the DCDM$\rightarrow$DR model, the perturbations of the DWDM species are identical to those of massive neutrinos~\cite{Ma:1995ey}. Moreover, the decay product perturbations are influenced by a collision term, which captures the fact that the DWDM species decays preferentially at low momenta~\cite{Holm:2022eqq}.

\begin{figure}[tb]
	\includegraphics[width=1.09\textwidth]{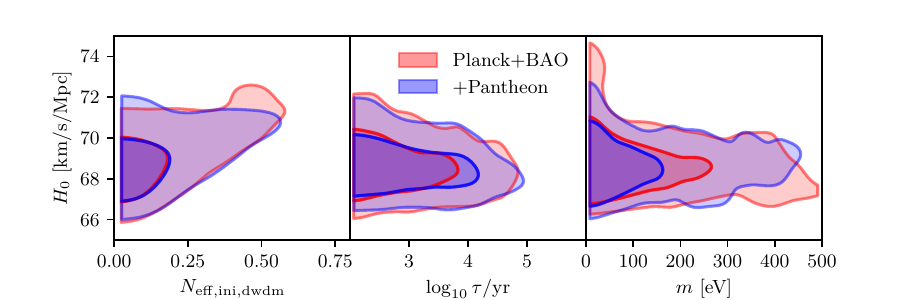}
	\caption{\textsl{Two-dimensional marginalized posteriors for the Hubble constant $H_0$ and the three parameters of the DWDM$\rightarrow$DR model: The initial abundance, $N_\mathrm{eff,ini,dwdm}$, parametrized as its contribution to the effective number of relativistic degrees of freedom, the DWDM lifetime, $\tau$, and the DWDM mass, $m$. Taken from Ref.~\cite{Holm:2022eqq}.}}
	\label{fig:dwdm}
\end{figure}

Figure~\ref{fig:dwdm}, adopted from~\cite{Holm:2022eqq}, shows contours of the two-dimensional marginalized posterior distributions for $H_0$ and three parameters of the DWDM$\rightarrow$DR model using Planck 2018 data combined with the full BOSS DR12 data set. Here, the initial abundance is parametrized as $N_\mathrm{eff,ini,dwdm}$, its contribution to the effective number of relativistic degrees of freedom. The prior on the lifetime studied here is roughly equivalent to the short-lived regime explored earlier.

Evidently, although the maximum a posteriori estimate favors the $\Lambda$CDM limit, a considerable amount of the DWDM species is permitted by the data. Ref.~\cite{Holm:2022eqq} finds a $68 \%$ credible upper bound of $N_\mathrm{eff,ini,dwdm} < 1.05$. Furthermore, as seen in figure~\ref{fig:dwdm}, there is a mild correlation between $H_0$ and the DWDM abundance, indicating that the DWDM$\rightarrow$DR model may admit a larger value of $H_0$ than $\Lambda$CDM. A $68 \%$ credible interval of $H_0 = 68.73^{+0.81}_{-1.3}\; {\rm km}\; {\rm s}^{-1}\; {\rm Mpc}^{-1}$, at a $2.7\sigma$ Gaussian tension with the representative local value $H_0 = 73.2 \pm 1.3\; {\rm km}\; {\rm s}^{-1}\; {\rm Mpc}^{-1}$~\cite{Riess:2020fzl}, is presented in this study. Thus, the broad conclusion is that there is no evidence from CMB data that the DWDM$\rightarrow$DR model solves the Hubble tension.

Although the constraints on the lifetime $\tau$ and mass $m$ are strongly driven by the bounds of the uniform prior chosen, in broad terms, the data prefers small masses and small lifetimes. Unfortunately, in all treatments of the DWDM$\rightarrow$DR model to date, the effects of \textit{inverse decays} have been neglected. The inverse decay process is kinematically allowed when the DWDM species decays while still relativistic, which is exactly the scenario in the preferred region of parameter space~\cite{Holm:2022eqq}. Thus, a complete understanding of the DWDM$\rightarrow$DR model inevitably requires a numerical implementation of the inverse decay processes and their quantum statistical corrections~\cite{Barenboim:2020vrr}. Finally, we also note that the Bayesian constraints on the DWDM$\rightarrow$DR model are expected to be strongly influenced by prior volume effects~\cite{Holm:2022eqq} since it reduces to the $\Lambda$CDM model in the limit of vanishing abundance, making the lifetime and mass unconstrained and thereby storing a large probability mass in the posterior distribution. At this time, a frequentist analysis of the model has not been conducted.

%%%%%%%%%%%%%%%%%%%%%%%%%%%%%%%%%%%%%%%%%%%%%%%%%%%%%%%%%

\section{Discussion and conclusion}

The three decay models presented in this chapter all show an ability to slightly alleviate the Hubble tension. The simplest of the models (DCDM$\rightarrow$DR), has been exhaustively studied, and it appears that it has reached its limits regarding how much it can alleviate the tension. The other two models, however, still have potential for further investigation. The DCDM$\rightarrow$DR+WDM model still needs a fast and accurate implementation of the full Boltzmann hierarchy, and with a fractional decay, this model can also be studied in the short-lived regime. The DWDM$\rightarrow$DR model also prefers slightly larger values of $H_0$, but a study including the inverse decay process is needed for a definite conclusion. Further investigation of the latter two models, along with improvements, could potentially alleviate the Hubble tension further. Ultimately, at the time of writing, we cannot definitively say which decaying dark matter model has the strongest alleviation of the Hubble tension, although there is a slight preference for the DWDM$\rightarrow$DR model with the $68\%$ C.I. $H_0 = 68.73^{+0.81}_{-1.3}\; {\rm km}\; {\rm s}^{-1}\; {\rm Mpc}^{-1}$ at a $2.7\sigma$ Gaussian tension with the representative local value $H_0 = 73.2 \pm 1.3\; {\rm km}\; {\rm s}^{-1}\; {\rm Mpc}^{-1}$~\cite{Riess:2020fzl}. Thus, although decaying dark matter models are somewhat more natural than several other proposed solutions to the $H_0$ discrepancy, they only accomplish a mild alleviation of the latter~\cite{Schoneberg:2021qvd}.

The next natural step to take in the hierarchy of numerical complexity would be a fully general two-body decay from a mother particle with a definite mass to two daughter particles with different masses, i.e., DWDM$\rightarrow$WDM$_{(1)}$+WDM$_{(2)}$. This is a challenging task since it combines all the most difficult aspects of the previous models while also introducing new ones. Although the full set of perturbation equations has been derived~\cite{Barenboim:2020vrr}, an implementation in a numerical Einstein--Boltzmann solver code still remains. Furthermore, we expect the full model to be very computationally expensive to evaluate, possibly making it unfeasible for immediate inference purposes. Nevertheless, a fully general decay scheme like this would, apart from other use cases like neutrino decays, possibly be able to affect the physics around recombination in just the right way for the Hubble tension to be relieved.

%%%%%%%%%%%%%%%%%%%%%%%%%%%%%%%%%%%%%%%%%%%%%%%%%%%%%%%%%

\begin{acknowledgement}
A.N., E.B.H., and T.T. were supported by a research grant (29337) from VILLUM FONDEN. We would like to thank Guillermo Franco Abellán (first author of Ref.~\cite{FrancoAbellan:2021sxk}) for letting us use figure~\ref{fig:abellan} in this chapter. 
\end{acknowledgement}

%%%%%%%%%%%%%%%%%%%%%%%%%%%%%%%%%%%%%%%%%%%%%%%%%%%%%%%%%

%\biblstarthook{References should be \textit{cited} in the text by number.\footnote{Please make sure that all references from the list are cited in the text. Those not cited should be moved to a separate \textit{Further Reading} section.} The reference list should be \textit{sorted} in alphabetical order. If there are several works by the same author, the following order should be used: 
%\begin{enumerate}
%\item all works by the author alone, ordered chronologically by year of publication
%\item all works by the author with a coauthor, ordered alphabetically by coauthor
%\item all works by the author with several coauthors, ordered chronologically by year of publication.
%\end{enumerate}
%For the reference style, we suggest to use \textit{LaTeX (US)} from INSPIRE.}

%\Emil{Vi skal ordne rækkefølgen af referencerne!}

\bibliographystyle{utcaps}
%\nocite{*}
\bibliography{bibliography}

%%%%%%%%%%%%%%%%%%%%%%%%%%%%%%%%%%%%%%%%%%%%%%%%%%%%%%%%%

\end{document}